\documentclass[useAMS,usenatbib]{mn2e}
\usepackage{graphicx}
\usepackage{amsmath}

\title[6668-MHz \& 6035-MHz in the Magellanic Clouds]{Multibeam Maser Survey of methanol and excited OH in the Magellanic Clouds: 
new detections and maser abundance estimates.}
\author[Green et al.]
       {J. A. Green$^1$\thanks{E-mail:james.green@postgrad.manchester.ac.uk}, J. L. Caswell$^2$, G. A. Fuller$^1$, S. L. Breen$^3$, K. Brooks$^2$, \newauthor
       M. G. Burton$^4$, A. Chrysostomou$^5$, J. Cox$^6$, P. J. Diamond$^1$, S. P. Ellingsen$^3$,  \newauthor
       M. D. Gray$^1$,  M. G. Hoare$^7$, M. R. W. Masheder$^8$, N. McClure-Griffiths$^2$, \newauthor
       M. Pestalozzi$^5$, C. Phillips$^2$, L. Quinn$^1$, M. A. Thompson$^5$, M. Voronkov$^2$, \newauthor
       A. Walsh$^{9}$, D. Ward-Thompson$^6$, D. Wong-McSweeney$^1$, J. A. Yates$^{10}$ \newauthor
        and R. J. Cohen$^{1}$\thanks{Deceased 2006 November 1.}
\\
$^1$ Jodrell Bank Centre for Astrophysics, Alan Turing Building, University of Manchester, Manchester, M13 9PL, UK; \\
$^2$ Australia Telescope National Facility, CSIRO, PO Box 76, Epping, NSW 2121, Australia; \\
$^3$ School of Mathematics and Physics, University of Tasmania, Private Bag 37, Hobart, TAS 7001, Australia; \\
$^4$ School of Physics, University of New South Wales, Sydney, NSW 2052, Australia;\\ 
$^5$ Centre for Astrophysics Research, Science and Technology Research Institute, University of Hertfordshire, College Lane, \\Hatfield, AL10 9AB, UK; \\ 
$^6$ Department of Physics and Astronomy, Cardiff University, 5 The Parade, Cardiff, CF24 3YB, UK; \\
$^7$ School of Physics and Astronomy, University of Leeds, Leeds, LS2 9JT, UK; \\ 
$^8$ Astrophysics Group, Department of Physics, Bristol University, Tyndall Avenue, Bristol, BS8 1TL, UK;  \\
$^{9}$ School of Maths, Physics and IT, James Cook University, Townsville, QLD 4811, Australia;\\
$^{10}$ University College London, Department of Physics and Astronomy, Gower Street, London, WC1E 6BT, UK
\\}

\date{Accepted XXXX . Received XXX; in original form XXXX}

\pagerange{\pageref{firstpage}--\pageref{lastpage}} \pubyear{XXXX}
\begin{document}
\maketitle

\label{firstpage}

\begin{abstract}
We present the results of the first complete survey of the Large and Small Magellanic Clouds 
for 6668-MHz methanol and 6035-MHz excited-state hydroxyl masers. In addition to 
the survey, higher-sensitivity targeted searches towards known star-formation regions 
were conducted. The observations yielded the discovery of a fourth 6668-MHz methanol 
maser in the Large Magellanic Cloud (LMC), found towards the star-forming region 
N160a, and a second 6035-MHz excited-state hydroxyl maser, found towards N157a. We 
have also re-observed the three previously known 6668-MHz methanol masers and the 
single 6035-MHz hydroxyl maser. We failed to detect emission from either transition in
the Small Magellanic Cloud. All observations were initially made using the Methanol 
Multibeam (MMB) survey receiver on the 64-m Parkes telescope as part of the MMB project 
and accurate positions have been measured with the Australia Telescope Compact Array  
(ATCA). We compare the maser populations in the Magellanic Clouds with those of our 
Galaxy and discuss their implications for the relative rates of massive star-formation, heavy 
metal abundance, and the abundance of complex molecules.
The LMC maser populations are demonstrated to be smaller than their Milky Way
counterparts. Methanol masers are under-abundant by a factor of $\sim$45,
whilst hydroxyl and water masers are a factor of $\sim$10 less
abundant than our Galaxy.
\end{abstract}


\section{Introduction}

Maser emission from star-forming regions in the Large Magellanic Cloud
(LMC) has been detected from a number of molecular transitions.
Initially, ground-state hydroxyl (OH) and water (H$_{2}$O) maser
transitions were detected (Caswell \& Haynes 1981; Haynes \& Caswell
1981; Scalise \& Braz 1982; Whiteoak et al. 1983; Whiteoak \& Gardner
1986) and more recently methanol (CH$_{3}$OH) maser emission at
6668 MHz (Sinclair et al. 1992; Ellingsen et al. 1994; Beasley et al.
1996). In all cases the detection rates and strength of maser emission
are lower than that seen in our own Galaxy.  

Before the present survey, only three 6668-MHz methanol masers were known to exist outside the
Milky Way, and all lay within the LMC. The first to be detected
was towards the nebula N105a/MC23 (Sinclair et al. 1992).  A second
was detected towards N11/MC18 as part of a targeted search of 48 
H{\sc ii} regions in the Magellanic Clouds (Ellingsen et al. 1994). A 
later search targeted 55 IRAS colour-selected sources and 12 known 
regions of bright H$\alpha$ emission and found a third 6668-MHz 
methanol maser toward IRAS 05011-6815 (Beasley et al. 1996). A 
recent, very sensitive, search of M33 failed to detect any further 
extragalactic methanol masers (Goldsmith, Pandian \& Deshpande 2007). 
The excited-state OH maser transition at 6035 MHz is often seen to be 
coincident with 6668-MHz methanol in our own Galaxy (e.g. Caswell 
1997), but only one extragalactic example has been reported; it is in the 
LMC, detected towards the nebula N160a/MC76 (Caswell 1995) and is 
not associated with any of the known methanol masers.

Searches of the SMC for 6668-MHz methanol masers by Ellingsen et al. 
(1994) towards 13 IRAS positions,  and by Beasley et al.  (1996) towards 
12 H{\sc ii} regions, failed to detect any emission. The only known 
masers in the SMC are 22-GHz H$_{2}$O masers at S7 and S9 (Scalise 
\& Braz 1982).

The SMC and LMC are both significantly less massive than the Milky
Way.  Stanimirovi\'c, Staveley-Smith \& Jones (2004) found a dynamical
H{\sc i} mass for the SMC of  2.4 $\times$ 10$^{9}$ M$_{\odot}$, 
while the mass of the LMC is between 6 $\times$ 10$^{9}$ and 1.5 
$\times$ 10$^{10}$ M$_{\odot}$ (Meatheringham et al. 1988; Schommer 
et al. 1992), compared to the Milky Way which has a mass of the order of 
5 $\times$ 10$^{11}$ M$_{\odot}$.  In addition the LMC has a star formation 
rate one tenth that of the Milky Way (Israel 
1980) and a lower metallicity, at 0.3 to 0.5 times the 
solar metallicity (Westerlund 1997). This will directly affect the ISM chemistry, 
and produce a lower dust-to-gas ratio, resulting in a higher ambient UV field
(Gordon et al. 2003).  Beasley et al. (1996) conclude that lower oxygen and 
carbon abundances in the LMC would lead to an under abundance of methanol 
by a factor of 6$-$12. Additionally, based on the work of Cragg et al. (1992), 
they believed, due to the significant role played by dust mid-IR emission in pumping this maser, that 
the lower dust abundance would contribute to fewer occurrences of methanol 
masers. The pumping model is now supported by even more accurate modelling (e.g. Cragg, Sobolev \& Godfrey 
2002; Cragg, Sobolev \& Godfrey 2005) and is widely accepted.

To provide a more complete and uniform view of the maser population in
the LMC and SMC, we report on the first complete systematic survey of the 
LMC and SMC made using the Parkes 64-m radio telescope. We present the 
results of the survey, supplemented with some deeper targeted Parkes 
observations, and some follow-up measurements with the Australia Telescope 
Compact Array (ATCA).

\section{Observations}
A new 7-beam 6$-$7 GHz receiver, jointly constructed by Jodrell Bank 
Observatory and the Australia Telescope National Facility (ATNF), is being used 
for the Methanol Multibeam (MMB) project described by Cohen et al. (2007). 
The principal aim of the project is to survey the entire Galactic plane, initially 
with the Parkes Radio Telescope in the southern sky, for both 6668-MHz 
methanol and 6035-MHz excited-state OH maser emission. In the sidereal time 
ranges when the Galactic plane is not visible from Parkes, both the LMC and the 
SMC are being surveyed.

Spectra for all MMB observations are taken with the multibeam and wideband correlators, 
yielding 2048 frequency channels across a bandwidth of 4~MHz, which corresponds to a velocity 
range of 180~km~s$^{-1}$ at 6668 MHz. The channel spacing is thus 1.95~kHz and corresponds to a 
velocity spacing of 0.09~km~s$^{-1}$ at 6668~MHz. The system temperature is typically 20 K (noise 
equivalent flux density of $<$60 Jy Hz$^{-1/2}$). The FWHM beam for methanol is 3.2 arcmin and for the 
excited-state OH it is 3.4 arcmin. The beams are arranged in an hexagonal pattern, spaced at 6.46 arcmin, 
so the total multibeam footprint is 15 arcmin wide. For further technical details on the MMB see Green 
et al. (in preparation).
  
Two techniques were employed to observe the LMC and SMC: the first was raster scanning of each galaxy 
as a whole, the second was targeted observations of known maser and star-formation regions. 

\begin{figure}
 \centering
\includegraphics[width=8.5cm]{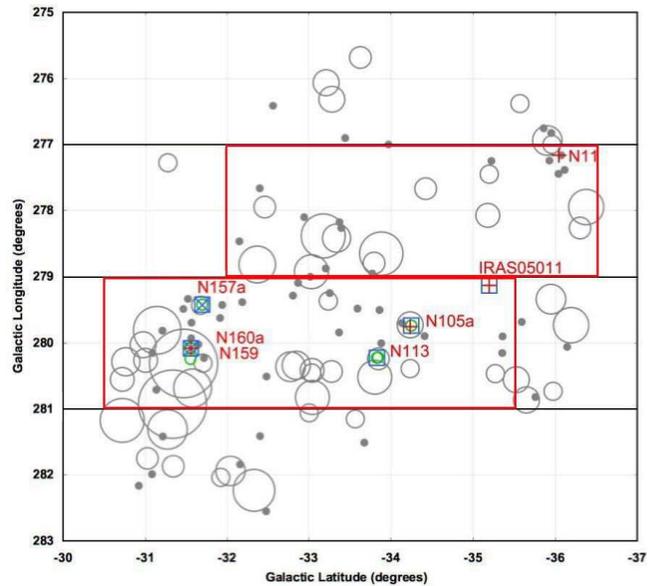}
\caption{\small Map of the survey region of the LMC with small and large CO (J = 1$-$0) clouds 
from the data of Fukui et al. (2001) with overlaid maser positions. Grey dots represent small CO
clouds and grey circles represent the large clouds, scaled proportionally to their total cloud surface area. 
Blue squares represent ground-state OH, green circles 22-GHz H$_{2}$O, blue crosses 6035-MHz excited-state OH, and red pluses 6668-MHz methanol.
The red solid lines encompass the higher sensitivity survey region.
}
\label{figureCO}
\end{figure}

\begin{figure}
 \centering
\includegraphics[width=8.5cm]{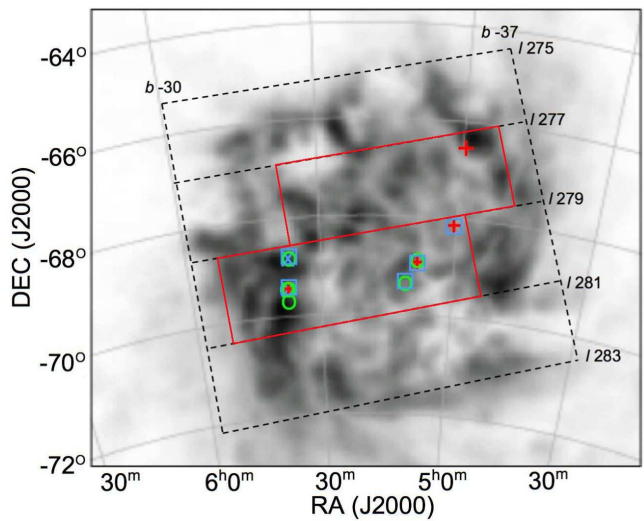}
\caption{\small Peak brightness-temperature image of H{\sc i} in the LMC (Staveley-Smith et al. 2003) overlaid with maser positions and Galactic coordinate blocks
used for scanning. The red solid lines encompass the higher sensitivity survey region.
The well known 30 Doradus nebula is at RA(J2000) 05$^{h}$38$^{m}$42.4$^{s}$, Dec.(J2000) -69$^{\circ}$06$'$02.8$''$.
Maser symbols are as per Fig \ref{figureCO}.}
\label{figureHI}
\end{figure}

Scanning was 
performed in Galactic coordinates, as this has the benefit of defining an almost rectangular region on the sky
and thus the scans are nearly parallel, and of similar length, over the whole region. The survey regions were 
chosen to fully sample the CO and H{\sc i} distributions of Fukui et al. (2001) and Staveley-Smith et al. (2003).
The LMC region surveyed is shown in Fig. \ref{figureCO}, along with the CO clouds contained within 
the region, and in Fig. \ref{figureHI} as an overlay on the H{\sc i} emission.
The LMC was scanned across 2$^{\circ}$ in longitude at a rate of 0.08$^{\circ}$ per min. These scans 
were separated by alternating latitude steps of 1.23 arcmin and 15 arcmin. This fully samples a 2$^{\circ}$ 
$\times$ 7$^{\circ}$ block of the LMC in 56 scans. The SMC was scanned across 3$^{\circ}$ in longitude 
at a rate of 0.15$^{\circ}$ per min, fully sampling a 3$^{\circ}$ $\times$ 4$^{\circ}$ block in 32 scans.

Data processing for the scanning technique is the same as for the MMB Galactic survey and is described 
in detail in Green et al. (in preparation). The scanning observations were made over the period 2006 
January to 2007 November, concurrently with the Galactic survey, making use of the complementary LST range. 
One pass was conducted over the full region, 275$^{\circ}$ $<$ 
$l$ $<$ 283$^{\circ}$, $-$30$^{\circ}$ $<$ $b$ $<$ $-$37$^{\circ}$ (four Galactic coordinate blocks of 56 
scans, each block requiring $\sim$20 hours observing time). This gave an rms noise of $\sim$0.22 Jy (which was poorer than subsequent passes).
All regions with detected CO (Fig. \ref{figureCO}) received a second pass and then 
as the 
known methanol masers were all within the middle two blocks (Fig. \ref{figureHI}), we chose to concentrate 
on these and thus conducted a further two passes over 277$^{\circ}$ $<$ $l$ $<$ 281$^{\circ}$, $-$30$^{\circ}$
$<$ $b$ $<$ $-$37$^{\circ}$, resulting in an rms noise of $\sim$0.09 Jy. We then further concentrated on the regions
 279$^{\circ}$ $<$ $l$ $<$ 281$^{\circ}$, $-$30.5$^{\circ}$ $<$ $b$ $<$ $-$35.5$^{\circ}$
and 277$^{\circ}$ $<$ $l$ $<$ 279$^{\circ}$, $-$32$^{\circ}$ $<$ $b$ $<$ $-$36.5$^{\circ}$.
These regions contain all the known masers, the largest CO clouds and over 70 per cent of the small CO clouds
(Fig. \ref{figureCO}).
A further four passes were conducted on these two regions (red outline in Fig. 1), resulting in an rms of  $\sim$0.06 Jy.
For the SMC we mapped 299$^{\circ}$ $<$ l $<$ 305$^{\circ}$, $-$42$^{\circ}$ $<$ b $<$ $-$46$^{\circ}$, 
taking two passes and resulting in an rms noise of $\sim$0.13 Jy. 

Targeted observations were performed toward 11 pointings in the LMC and two in the SMC (see Table 1).
For the LMC the targeted regions were N11, IRAS 05011-6815, N105a, N113, N157a, N159 and 
N160a, with two positions for N105a, N113, N157a and N160a. For the 
SMC the targeted regions were the positions of two known 22-GHz H$_{2}$O 
masers, S7 and S9 from Scalise \& Braz (1982). These points in the sky were tracked with the pointing centre 
cycling through each of the seven receiver beams. This means each of the seven receiver beams was on source for 
10 minutes (with the exception of N157a where 20 minute integrations were taken). Data processing for the 
targeted pointings used the package ASAP, developed by Chris Phillips and Malte Marquarding at the ATNF. 
The bandpass reference for each beam was estimated using the median spectrum of the 6  `off-source' positions, 
and each median reference then combined with the corresponding total power spectrum
`on-source' to determine a baseline- and gain-corrected quotient spectrum.
The seven spectra were then
combined to give a final best spectrum with an effective 70 min integration time (140 min for N157a) and a 
typical rms of $\sim$25 mJy in the total intensity spectrum.
Flux densities were calibrated using observations of the continuum source PKS 1934-638. 
The targeted Parkes observations were taken in 2006 June (N11, N105a, IRAS 05011-6815), 2006 September 
3 (N160a) and 2007 January 25 (N157a) and are listed with their rms noise levels in Table \ref{Parkestable}.

In addition to the targeted Parkes observations, the Australia Telescope Compact Array (ATCA) was used 2007 July 21$-$22
to obtain precise positions for the new methanol maser at position 11 and for the two excited-state OH masers at 
positions 8 and 11. These are listed in Table 2. A loop of the three targets and the phase calibrator 0530-727 was repeated 
22 times over a 10-hour time span and five times on the following day. Within each loop, the three targets had
integration times of 15, 2 and 4 minutes respectively.
The array was in a 6 km configuration (6C), yielding a FWHM beamsize of 2.8 $\times$ 
1.6 arcsec. The bandwidth was 4 MHz, spread across 2048 frequency channels, the same as for the Parkes spectra.  

\begin{table}
\begin{minipage}{80mm}
\small
\centering
\caption{\small Targeted regions: Parkes pointing positions and rms noise levels for the total intensity spectrum.}
\begin{tabular}{l l l c c}
\\
\hline
 \multicolumn{1}{l}{Name} & \multicolumn{1}{l}{Pos.} & \multicolumn{2}{c}{Pointing} &\multicolumn{1}{l}{rms}\\
 & & \multicolumn{1}{c}{RA(J2000)} & \multicolumn{1}{c}{Dec(J2000)} & \multicolumn{1}{l}{noise}  \\
 & & \multicolumn{1}{c}{$h$ $m$ $s$}  & \multicolumn{1}{c}{$^{\circ}$ $'$ $''$} &  (mJy) \\
\hline 
 N11/MC18 &1& 04 56 47 & $-$66 24 35  &  23 \\
 IRAS 05011&2 & 05 01 02  & $-$68 10 28 & 26 \\
 N105a/MC23 &3& 05 09 52 & $-$68 53 29 & 26 \\
&4 & 05 09 59  & $-$68 54 34 & 26 \\
N113/MC24 &5& 05 13 25 & $-$69 22 46  & 26 \\
 &6& 05 13 18 & $-$69 22 21 & 26 \\
 N157a/MC74 &7& 05 38 47  & $-$69 04 46  & 20 \\
 & 8 & 05 38 45  & $-$69 05 07 & 20 \\
 N159 &9& 05 39 29  & $-$69 47 19 &  26 \\
 N160a/MC76 &10& 05 39 44 & $-$69 38 34& 26 \\
 &11& 05 39 39  & $-$69 39 11  & 26 \\
 S7 & 12 & 00 46 39 & $-$72 40 49 & 22\\
 S9 & 13 & 00 47 31 & $-$73 08 20 & 21\\
\end{tabular} 
\label{Parkestable}
\end{minipage}
\end{table}

\section{Results}
\subsection{Scanned Observations}
This is the first time both the LMC and SMC have been mapped in their entirety 
for 6668-MHz methanol and 6035-MHz OH maser emission. In the LMC region 
of 277$^{\circ}$ $<$ $l$ $<$ 281$^{\circ}$, $-$30$^{\circ}$ $<$ $b$ $<$ 
$-$37$^{\circ}$, with rms noise of $\sim$0.09 Jy, we detected ($\ge$3$\sigma$): 
the known 6668-MHz sources at IRAS 05011-6815 and N11;  the known 6035-MHz 
OH source at N160a; and marginally ($\sim$2$\sigma$) detected the known 
6668-MHz source at N105a. 

In the SMC, with rms noise of 0.13 Jy, we failed to detect any maser emission from 
either the 6035-MHz OH or the 6668-MHz methanol transitions. 

\subsection{Targeted Observations}
\begin{figure}
 \centering
\includegraphics[width=7.5cm]{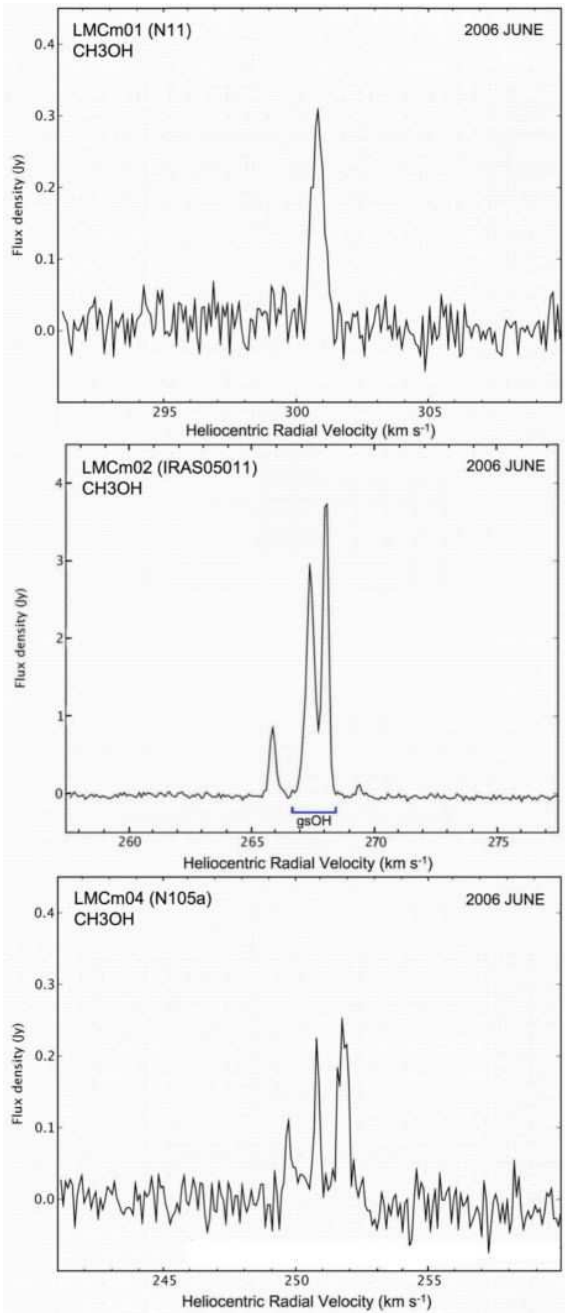}
\caption{\small 6668-MHz methanol maser sources in the LMC known prior to the current study. 
Spectra from the Parkes telescope. From top to bottom: LMCm01, LMCm02, LMCm04. The horizontal 
line on the spectrum of LMCm02 represents the velocity range of the ground-state OH (gsOH) also found
at the same site.}
\label{figure1}
\end{figure}

\begin{figure}
\centering
\includegraphics[width=7.5cm]{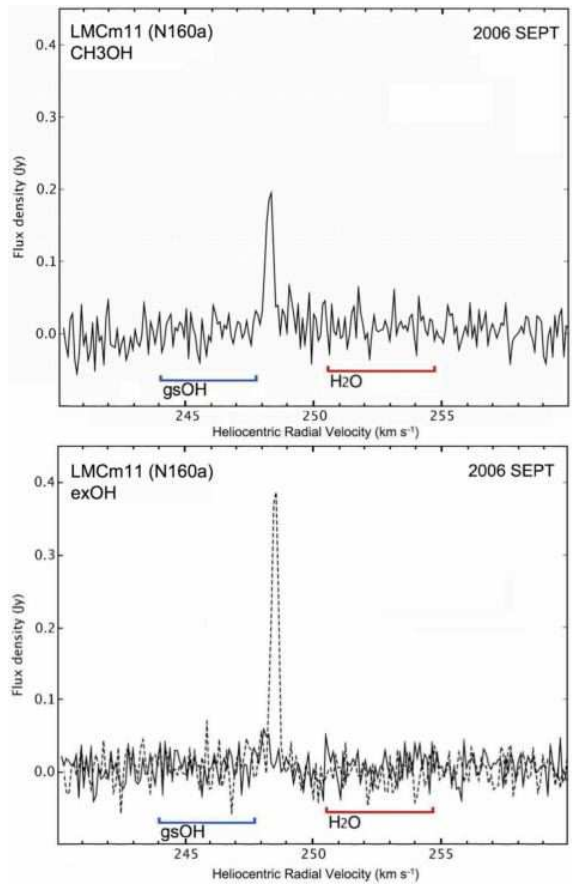}
\caption{\small The newly detected 6668-MHz methanol maser spectra (top) 
and known 6035-MHz excited-state OH spectra (bottom). For the excited-state OH, 
the dashed line is LHC polarization, solid line is RHC polarization, and the flux density 
scale refers to a single polarization. Horizontal lines represent the velocity range of the 
other known masers, H$_{2}$O and ground-state OH (gsOH).}
\label{figure2}
\end{figure}

\begin{figure}
 \centering
\includegraphics[width=7.5cm]{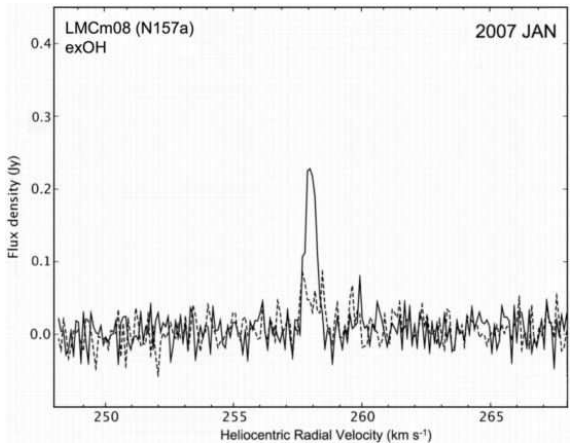}
\caption{\small Parkes telescope spectra of the new 6035-MHz OH maser source detected 
toward N157a in the LMC. The dashed line is LHC polarization, solid line is RHC polarization, 
and the flux density scale refers to a single polarization.}
\label{figure3}
\end{figure}

The results of the targeted observations are given in Table \ref{table1}. In addition 
to detecting the three known methanol masers at N11, N105a and IRAS 05011-6815 
(Fig. \ref{figure1}), a new methanol maser was detected ($>$5$\sigma$) in association 
with the known excited-state OH maser in N160a (Fig. \ref{figure2}). A new excited-state 
OH maser was also detected ($>$5$\sigma$), through a pointing at N157a (Fig. 
\ref{figure3}). Nothing was detected towards the SMC pointings (Table 1), nor towards
a recently discovered H$_{2}$O maser (J. Lovell, private communication). 

The two new sources were, with hindsight, marginally ($\sim$3$\sigma$) detected in the 
survey data. Each pointing is discussed in detail in the next Section.

\subsection{Comment on Individual Regions}
In Table 2 we summarize for the LMC our new data on six masers of methanol and excited-state OH (at five locations)
and complementary ground-state OH and water data. In addition, for completeness, we list from the literature the 
other six reliably positioned locations hosting masers of ground-state OH and water. 

The additional data are from Brooks \& Whiteoak (1997) for OH, and from Lazendic et al. (2002) for water. 
A more recent search for water by Oliveira et al. (2006) found no distinct new maser locations. The positional
errors cited in Table 2 differ in some cases from the original references since we re-assessed those that seemed not
to incorporate realistic systematic errors. We discuss the 11 sites individually so as to highlight the associations
between different maser species. We have introduced a new designation for the maser positions so that each
site can be easily referred to unambiguously. We use the prefix 'LMCm' to denote LMC maser, and follow
this with a sequence of numbers running from 01 to 11. Various species at the same site are distinguished by specifying
the species rather than referring to them by a different number.

\subsubsection{N11/MC18 (LMCm01: methanol only)}
N11 is the second largest H{\sc ii} region in the LMC, located at the southern edge of the LMC-1 supergiant shell, 
and ionized through OB association hot stars. No maser emission 
from ground-state OH or 22-GHz H$_{2}$O has been detected to date. However, N11 was found to show 6668-MHz methanol 
maser emission by Ellingsen et al. (1994) with a peak flux density of 0.3 Jy at a heliocentric 
velocity of 301 km s$^{-1}$, at the position given in Table 2.  The current observations detect this methanol 
maser at the same velocity with a peak flux density of 0.33 Jy, but failed to detect any 6035-MHz OH maser 
emission. Comparing the methanol spectra of the existing observations with the new implies that any
variations over the 12 year separation in observations are below the uncertainty of the measurements.

\subsubsection{IRAS 05011-6815 (LMCm02: methanol and ground-state OH)}
IRAS 05011-6815 meets the Wood \& Churchwell (1989) colour criteria for an ultra-compact (UC) H{\sc ii} region, 
although it does not exhibit any 1.6 GHz or 8.8 GHz continuum emission (Brooks \& Whiteoak 1997; 
Beasley et al. 1996). Beasley et al. (1996) searched for 6668-MHz methanol emission based on the colour
criteria for the IRAS source and found three components with 265.6 $<$ V$_{\rm helio}$ $<$ 267.9 km s$^{-1}$ 
with a position as given in Table 2. Brooks \& Whiteoak (1997) detected maser emission at both 1665-MHz and 
1667-MHz transitions of OH. The 1665-MHz OH exhibited a peak flux density of 0.23 Jy beam$^{-1}$ at a 
heliocentric velocity of 267.6 km s$^{-1}$, whilst the 1667-MHz OH had a peak flux density of 0.13 Jy beam$^{-1}$ 
at 267.6 km s$^{-1}$. No 22-GHz H$_{2}$O maser emission has been detected towards the source. As noted by 
Brooks \& Whiteoak (1997) the positions of the 1665 and 1667 MHz OH are coincident with the 6668-MHz 
methanol to within their errors. The methanol observations presented here find three clear components that correspond 
with those found previously and one new component at V$_{\rm helio}$ of 269.1 km s$^{-1}$. The four components 
have peak flux densities of 0.89, 2.96, 3.78 and 0.11 Jy respectively. The three known components are comparable
to those that Beasley et al. (1996) measured with the Hobart antenna. However, the ATCA flux densities found by Beasley et al. were lower, 
perhaps due to poorer spectral resolution. 

\subsubsection{N105a/MC23 (LMCm03: ground-state OH and water; LMCm04: methanol only)}
N105a is an H{\sc ii} region with 1.6-GHz and 6-GHz continuum emission (Brooks \& Whiteoak 1997; 
Ellingsen et al. 1994).  Brooks \& Whiteoak (1997) determined an accurate position for ground-state OH maser emission, not only 
at 1665-MHz (previously observed by both Haynes \& Caswell 1981 and by Gardner \& Whiteoak 1985), 
but also for newly detected emission at 1667-MHz. The 1665-MHz OH had a main component with a peak 
flux density of 0.58 Jy beam$^{-1}$ seen at a velocity of  253.4 km s$^{-1}$ and a weak (0.09 Jy 
beam$^{-1}$) component at 255.8 km s$^{-1}$.  The 1667-MHz OH had one component at 254.2 km 
s$^{-1}$ with a peak flux density of 0.25 Jy beam$^{-1}$. An average OH position is given in Table 2.
 H$_{2}$O maser emission at 22.2 GHz was 
detected by Scalise \& Braz (1981), Whiteoak et al. (1983), Whiteoak \& Gardner (1986) and Lazendic 
et al. (2002), and found to have a number of features between 247 km s$^{-1}$ and 273 km s$^{-1}$, 
with flux densities ranging from 0.8 to 1.8 Jy. The average position of these features is coincident 
with the ground-state OH to within the errors.

Sinclair et al. (1992) first observed methanol maser emission from N105a with a total intensity of 0.12 Jy 
and two features at heliocentric velocities of 250.7 km s$^{-1}$ and 251.5 km s$^{-1}$. N105a was then 
re-observed by Ellingsen et al. (1994) and the peak flux density was found to be 0.17 Jy with one of the 
features at a slightly different heliocentric peak velocity (the second was at 251.8 km s$^{-1}$ rather than 
251.5 km s$^{-1}$). They also found a potential weaker (0.1 Jy) feature at 249.7 km s$^{-1}$. The current 
observations exhibit all three components, thus confirming the weaker feature. The three have heliocentric 
velocities of 249.8, 250.8 and 252.0 km s$^{-1}$ and respective flux densities of 0.11, 0.22 and 0.25 Jy. 
These flux densities imply a slight increase over those of Ellingsen et al. (1994), but the spectral shapes are 
consistent. Brooks \& Whiteoak measured the position of the methanol maser in 1997 with the ATCA and this 
is the position listed in Table 2. We made additional observations 2007 November and confirmed this position
to better than 1 arcsec.

\begin{table*}
\begin{minipage}{180mm}
\small
\centering
\caption{\small  Four 6668-MHz methanol masers and two 6035-MHz OH masers in the LMC with additional data on other
species. Peak flux densities (S$_{\rm peak}$) are those recorded at Parkes on 2006 June 4 (N11), 2006 June 6 (N105a), 2006 June 7
(IRAS 05011-6815), 2006 September 3 (N160a), and 2007 January 25 (N157a). Positions (with rms errors in parentheses) 
are from ATCA measurements as discussed in the source notes of Section 3.3. $^{\dagger}$Other known transitions are listed
together as a single position when they coincide to within the uncertainties. They are referenced as:
$^{1}$Caswell \& Haynes 1981, $^{2}$Haynes \& Caswell 1981,  $^{3}$Scalise \& Braz 1982, $^{4}$Whiteoak et al. 1983,  
$^{5}$Gardner \& Whiteoak 1985, $^{6}$Whiteoak \& Gardner 1986, $^{7}$Brooks \& Whiteoak 1997, $^{8}$Lazendic et al. 2002.}
\begin{tabular}{c c c c c c c l}
\\
\hline
Maser & \multicolumn{1}{c}{RA} & \multicolumn{1}{c}{Dec} &\multicolumn{2}{l}{6668-MHz Methanol} 
& \multicolumn{2}{l}{6035-MHz OH} & \multicolumn{1}{l}{Other known transitions$^{\dagger}$} \\
Site & (J2000) & (J2000) & \multicolumn{1}{c}{V$_{\rm helio}$} & \multicolumn{1}{c}{S$_{\rm peak}$}   
& \multicolumn{1}{c}{V$_{\rm helio}$} & \multicolumn{1}{c}{S$_{\rm peak}$} & \\
Number & $h$ $m$ $s$ & $^{\circ}$ $'$ $''$ & (km s$^{-1}$) & (Jy) & (km s$^{-1}$) & (Jy) &  \\
\hline 
LMCm01 & 04 56 47.10 (0.03) & $-$66 24 31.7 (0.2) & 301.0 & 0.33 & $-$ & $-$ & $-$ \\
LMCm02 & 05 01 01.85 (0.06) & $-$68 10 28.3 (0.3) & 268.0 & 3.78 & $-$ & $-$ & 1665/1667-MHz OH$^{7}$ \\
LMCm03 & 05 09 52.00 (0.04) & $-$68 53 28.6 (0.3) & $-$ & $-$ & $-$ & $-$ & 1665/1667-MHz OH$^{2,5,7}$ \\
&&&&&&&  22-GHz H$_{2}$O$^{4,6,8}$ \\
LMCm04 & 05 09 58.66 (0.03) & $-$68 54 34.1 (0.2) & 252.0 & 0.25 & $-$ & $-$ & $-$ \\
LMCm05 & 05 13 25.18 (0.07) & $-$69 22 46.0 (0.4) & $-$ & $-$ & $-$ & $-$ & 1665-MHz OH$^{7}$ \\
&&&&&&& 22-GHz H$_{2}$O$^{6,8}$  \\
LMCm06 & 05 13 17.67 (0.07) & $-$69 22 21.4 (0.4) & $-$ & $-$ & $-$ & $-$ & 22-GHz H$_{2}$O$^{8}$ \\
LMCm07 & 05 38 46.65 (0.07) & $-$69 04 45.5 (0.4) & $-$ & $-$ & $-$ & $-$ &  22-GHz H$_{2}$O$^{4,6,8}$ \\
LMCm08 & 05 38 45.00 (0.03) & $-$69 05 07.4 (0.2) & $-$ & $-$ & 258.0 & 0.22 & $-$\\
LMCm09 & 05 39 29.41 (0.07) & $-$69 47 18.9 (0.4) &  $-$ &  $-$ &  $-$ &  $-$ & 22-GHz H$_{2}$O$^{3,8}$ \\
LMCm10 & 05 39 43.92 (0.10) & $-$69 38 33.6 (0.6)& $-$ & $-$  & $-$ & $-$ & 22-GHz H$_{2}$O$^{6,8}$\\
LMCm11 & 05 39 38.94 (0.03) & $-$69 39 10.8 (0.2) & 248.0 & 0.20 & 248.5 & 0.39 & 1665-MHz OH$^{1,5,7}$, \\
&&&&&&& 22-GHz H$_{2}$O$^{4,6,8}$\\
\end{tabular} 
\label{table1}
\end{minipage}
\end{table*}

\subsubsection{N113/MC24 (LMCm05: ground-state OH and water; LMCm06: water only)}
N113 is an H{\sc ii} region with 1.6-GHz continuum emission (Brooks \& Whiteoak 1997). Brooks \& Whiteoak 
detected only the 1665-MHz transition of the ground-state OH maser in N113, with the position listed in
Table 2. This was at a velocity of 248.3 km s$^{-1}$ with a peak flux density of 0.26 Jy beam$^{-1}$.  
Water maser emission at 22-GHz was observed by Lazendic et al. (2002) towards two locations, 
the first of which was initially observed by Whiteoak \& Gardner (1986). One site coincides with the 
ground-state OH, the other is found alone at position LMCm06 of Table 2. The maser emission was seen across 
velocity ranges of 248 to 258 km s$^{-1}$ and 249 to 252 km s$^{-1}$, with peak flux densities of 82 Jy and 
3.2 Jy, respectively. The current study failed to detect either 6668-MHz methanol or 6035-MHz 
excited-state OH maser emission.

\subsubsection{N157a/MC74 (LMCm07: water only; LMCm08: excited-state OH only)}
N157a is essentially the 30 Doradus region of star-formation, surrounding the star cluster NGC2070 and 
is one of the most active starburst regions  in the local group. No ground-state 1665/1667-MHz OH emission
has been detected from this region (Brooks \& Whiteoak 1997). Water maser emission at 22-GHz is seen 
at a peak heliocentric velocity of 269.7 km s$^{-1}$ with a peak flux density that varied between 2.5 and 4.5 Jy for
a sequence of observations between 1982 and 1984 (Whiteoak et al. 1983; Whiteoak \& Gardner 1986). It was 
then detected with a peak flux density of 3.7 Jy (Lazendic et al. 2002) at a position as given in Table 2, position LMCm07.
Our targeted search failed to detect any 6668-MHz methanol emission, but did discover a new site of 
6035-MHz OH emission with a peak flux density of 0.22 Jy at a heliocentric velocity of 258 km s$^{-1}$, 
and not spatially coincident with the water maser. Emission at 6035-MHz in the absence of 1665-MHz emission 
is rare (Caswell 2004). In this instance, we note that 1665 and 1667 MHz absorption has been detected towards 
MRC0539-691 with the wide (12 arcmin) beam of the Parkes telescope (Gardner \& Whiteoak 1985);  it was 
not detected with the 7 arcsec beam of the ATCA (Brooks \& Whiteoak, 1997), but may, never the less be responsible for 
either masking or even absorbing any maser emission at this position.

\subsubsection{N159 (LMCm09: water only)}
N159 is an H{\sc ii} region with 1.6-GHz continuum emission (Brooks \& Whiteoak 1997). Brooks \& Whiteoak 
(1997) failed to detect any ground-state OH maser emission. N159 exhibited 22-GHz H$_{2}$O maser emission 
in 1981 (Scalise \& Braz 1982), but was not detected by Whiteoak \& Gardner in 1986. It was then re-observed 
by Lazendic et al. (2002) with a peak flux density of 3.7 Jy at a heliocentric velocity of 247.5 km s$^{-1}$
allowing the precise position measurement given as position LMCm09 of Table 2. Sinclair et al. (1992) failed to detect any 
6668-MHz methanol at a sensitivity of 0.03 Jy rms in 1991 and the current study also failed to detect emission
in either 6668-MHz methanol or 6035-MHz OH.

\begin{table*}
\begin{minipage}{180mm}
\small
\centering
\caption{\small Comparison of Maser Populations. References: $^{a}$Caswell \& Haynes (1987), 
$^{b}$ Greenhill et al. (1990)
 $^{c}$Sinclair et al. (1992); $^{d}$Ellingsen et al. (1994); $^{e}$Caswell \& Vaile (1995), $^{f}$Beasley et al. (1996), 
 $^{g}$Brooks \& Whiteoak (1997), 
 $^{h}$Lazendic et al. (2001) $^{i}$Pestalozzi et al. (2005) \& Pestalozzi et al. (2007); 
 $^{j}$Current Work,  $^{k}$MMB;  $^{\dagger}$Although the Galactic ground-state OH population is unlikely
to change significantly, both the methanol and excited-state OH Galactic populations will be better defined once the 
MMB is completed and the excited-state OH Galactic population could conceivably increase. Therefore the 1:1 ratio for excited-state OH 
should be considered with this in mind. Luminosities for the LMC are based on a distance of 50 kpc and Galactic 
luminosities assume that the Galactic centre is 8.5 kpc from the Sun. $^{*}$See section 4.1.2.}
\begin{tabular}{l c c l l c}
\\
\hline
Molecule & \multicolumn{1}{c}{Transition} & \multicolumn{1}{c}{Luminosity Cutoff} &\multicolumn{1}{l}{LMC Pop.} 
& \multicolumn{1}{l}{Galactic Pop.} & \multicolumn{1}{c}{Pop. Ratio} \\
& (MHz) & (Jy kpc$^{2}$) & & & \multicolumn{1}{c}{(Gal/LMC)} \\
\hline 
OH & 1665 & 500 & $\ge$4$^{g}$ & 60$^{a}$ & $\le$15 \\
OH & 6035 & 500 & $\ge$2$^{e,j}$ & 2$^{e}$ & $\le$1$^{\dagger}$,$\le$10$^{*}$ \\
CH$_{3}$OH & 6668 & 500 & $\ge$4$^{c,d,f,j}$ & 186$^{i,k}$ & $\le$47$^{\dagger}$ \\
H$_{2}$O & 22235 & 2500 & $\ge$7$^{h}$ & 88$^{b}$ & $\le$13 \\
\end{tabular} 
\label{table2}
\end{minipage}
\end{table*}

\subsubsection{N160a/MC76 (LMCm10: water only; LMCm11: ground-state OH, excited-state OH, methanol and water)}
N160a is an H{\sc ii} region with 6-GHz continuum emission (Caswell 1995). OH maser emission at 1665-MHz is 
seen at 248.6 km s$^{-1}$ with a peak flux density of 0.22 Jy (Caswell \& Haynes 1981; 
Gardner \& Whiteoak 1985; Brooks \& Whiteoak 1997). A second feature at the same position (position LMCm11 of Table 2) was also observed 
with a flux density of 0.08 Jy at 244.5 km s$^{-1}$ (Brooks \& Whiteoak 1997). No 1667-MHz OH 
maser emission has been detected to date. H$_{2}$O maser emission at 22-GHz is found at two sites in N160a, one,
at position LMCm11, is a narrow feature at 252.9 km s$^{-1}$ with a peak flux density of 5.1 Jy (Lazendic et al. 2002). 
The other covers a range of velocities from 245 to 261 km s$^{-1}$, peaking at 253 km s$^{-1}$ with a flux density of 1.6 Jy 
(Whiteoak \& Gardner 1986; Lazendic et al. 2002). This is listed as position LMCm10. 
The current observations detect 6035-MHz excited-state OH maser 
emission, which was first seen by Caswell (1995). The Caswell spectra, obtained in 1994, showed three RHC components 
and two LHC components with LSR velocities between 243.4 and 249.5 km s$^{-1}$ and peak flux densities between 
0.06 and 0.41 Jy. The previous position was compatible with our improved estimate in Table 2. 
The current observations show only the strong LHC component at 248.5 km s$^{-1}$, which has a peak flux density of 0.39 Jy
(similar to its 1994 value) and possibly the RHC feature at 248.0 km s$^{-1}$. Other components have weakened below our 
detection threshold.  
The current survey detected 6668-MHz methanol maser emission for the first time 
in this region. The maser emission was at a heliocentric velocity of 248.0 km s$^{-1}$ with a peak flux density of 0.20 Jy.
The ground-state OH, the excited-state OH and the 6668-MHz methanol are all spatially coincident to within the errors,
with excited-state OH providing the most precise position.

\section{Discussion}
\subsection{Maser Abundance in the Magellanic Clouds}
Masers are a potentially important tracer of star-formation in galaxies, but 
how the maser emission is affected by environmental parameters such as 
metallicity and the ambient UV field is largely unknown. Comparison of the 
LMC, SMC and Milky Way maser populations can start to provide insight into 
this. We summarise the relevant results in Table 3. Ideally this comparison would be done with the Galactic luminosity 
distribution of each maser species, as estimated with varying degrees of 
confidence in the following sections. The luminosity parameter that we adopt is 
simply the peak flux density multiplied by the square of the distance. We assume 
the luminosity to be isotropic. For the LMC we adopt a distance of 50 kpc and 
for the SMC 60 kpc (Feast 1999; Walker 1999). The lowest luminosity 
maser observed in the Magellanic Clouds for each species is taken as a `cutoff', and 
for sources above this cutoff an empirical ratio of the size of the maser population 
can be estimated, as given in the final column. The Galactic populations are taken 
as the estimated number of masers above the equivalent Galactic `cutoff'. In cases 
where the Galactic population is well understood, but the Magellanic Cloud 
population is incomplete (even above the cutoff), then the ratio will be an upper 
limit.

\subsubsection{22-GHz H$_{2}$O}
For 22-GHz H$_{2}$O we revisit the discussion of Brunthaler et al. (2006),
and note from Lazendic et al. (2002) that the weakest peak flux density of
this species of maser in the LMC is 1.6 Jy. The derived Galactic luminosity
function of Greenhill et al. (1990) is based on maser linewidths of 1 km 
s$^{-1}$, whereas the weakest LMC detections are narrower than this.  So, 
in contrast to Brunthaler et al. who adopted linewidths of 2.1 km s$^{-1}$, 
we have smoothed the LMC luminosities to linewidths of 1 km s$^{-1}$, and 
hence arrive at a threshold of $\sim$1 Jy. Therefore we adopt 2500 Jy 
kpc$^{2}$ as the luminosity threshold (a factor of $\sim$5 greater than the 
value of Brunthaler et al.). Then, using the same luminosity function of 
Greenhill et al. (1990) we find the Galactic population corresponds to 88
and this gives a ratio of $\sim$13. This is subject to the caveat that both 
the Galactic and Magellanic populations are likely to be severely incomplete, 
principally due to small beams at this frequency making large-scale searches 
difficult.

\subsubsection{1665-MHz and 6035-MHz OH}
The ground-state OH luminosity function in the desired form is provided by 
Caswell \& Haynes (1987). However, their luminosity scaling assumes a 
Sun$-$Galactic Centre distance of 10 kpc, rather than the currently preferred 
8.5 kpc, so we adjust their luminosity scale accordingly. The Brooks \& 
Whiteoak (1997) measurements list four sites of maser emission in the LMC, 
with strongest peaks ranging from 221 mJy to 580 mJy. Their measurements 
could have detected sources as weak as 50 mJy in favourable cases, but we treat 
200 mJy as the cutoff, i.e. 500 Jy kpc$^{2}$. The results do not represent a 
complete survey of the LMC, so four is a lower limit to the population above this 
threshold. From the Caswell \& Haynes (1987) estimate of the Galactic population, 
we estimate that 60 sources lie above the corrected cutoff of 500 Jy kpc$^{2}$, 
therefore the ratio is $\le$15.

For the 6035-MHz transition of OH, the brighter polarization of the source in N157a 
has a luminosity of $\sim$625 Jy kpc$^{2}$, and the source in N160a has an even 
higher luminosity of $\sim$1000 Jy kpc$^{2}$. Hence, for this transition also, we 
adopt an approximate cutoff of 500 Jy kpc$^{2}$. From the discussion of Caswell \& 
Vaile (1995), updated to include information from Caswell (2001; 2003), we find only 
two sources in the Galaxy to be above this threshold. At face value it is quite surprising 
to find the LMC population apparently equalling that of the Galaxy. However, these are 
clearly very small number statistics, and we may attempt to counter this problem by using
information on the whole Galactic population (exceeding 100). A general comparison with 
ground-state OH suggests that, to similar luminosity limits, the 6035-MHz masers are 
one-third as common (Caswell 2001). That would suggest a Galactic population of 20 above 
our cutoff, and a ratio of $\sim$10.

\subsubsection{6668-MHz Methanol}
For the 6668-MHz methanol masers, the source at IRAS 05011-6815, peaking just shy of 4 Jy in the LMC, 
corresponds to a luminosity of $\sim$10000 Jy kpc$^{2}$ and the sources in N11, N105a and N160a would 
be between $\sim$500 and $\sim$750 Jy kpc$^{2}$. This gives us a luminosity threshold of 500 Jy kpc$^{2}$. 
For the the Galactic population, Pestalozzi et al. (2007) analyze the Pestalozzi et al. (2005) compilation of 
6668-MHz methanol masers and have evaluated the luminosities assuming all sources to be at the near 
kinematic distance (Fig. 7 of Pestalozzi et al. 2007). Converting their solar luminosities to our units, and 
summing the number of sources above 500 Jy kpc$^{2}$, we find 169 sources above the cutoff. If some 
sources are at the far distance, then the number would be increased;  Pestalozzi et al (2007) argue that most 
of the sources are indeed at the near distance, so we assume the correction to be small and ignore it for the 
present.  We do however make a correction for incompleteness in the catalogue, since analysis of the MMB
Galactic survey to date (Cohen et al. 2007; Green et al. 2007) indicates that the number of sources in the high 
flux tail of the Pestalozzi et al. distribution is probably underestimated by about 10 per cent. So for the purposes 
of comparison we have corrected for this additional factor (increasing the Galactic population above the threshold 
to 186). This results in a difference of populations by a factor of $\sim$45. As the methanol results are based on 
a complete survey of the LMC, the population ratio for the maser species should be the most reliable. It therefore 
seems significant that, compared to our own Galaxy, the methanol masers in the LMC are under-abundant relative 
to OH. This is perhaps suggestive that the LMC environment is less conducive for methanol emission, than OH.

\subsubsection{Factors affecting the detection rates}
The extragalactic maser populations are likely to reflect the star-formation rates 
of their parent galaxies. Israel (1980) estimated the rates for the LMC, SMC and 
Milky Way as 0.4 M$_{\odot}$ yr$^{-1}$, 0.08 M$_{\odot}$ yr$^{-1}$ and 
3.5 M$_{\odot}$ yr$^{-1}$ respectively. For the LMC it should be noted that 
the 30 Doradus complex contributes almost a third of the flux used to calculate 
the star formation rate, and so it may not be a good measure of the typical star 
formation rate throughout the galaxy. However, at face value these rates imply 
the LMC should have $\sim$9 times fewer masers, the SMC $\sim$40 times 
fewer than the Milky Way. The ground-state OH, and perhaps the water masers, 
appear to be broadly compatible with these expectations, but the methanol 
population appears to be four or five times smaller than expected. For the Israel 
rates the timescale for comparison was the mean lifetime of Lyman continuum 
producing stars, equivalent to the lifetime of a O~6.5 star (5 $\times$ 10$^{6}$ 
years). In contrast, the lifetime for masers is believed to be $<$ 10$^{5}$ years 
(Forster \& Caswell 1989); therefore the masers only trace recent star formation, 
which may well differ from the mean rate over 5$\times$10$^{6}$ years.

However both the mass of the forming stars, and their environment, will affect 
the incidence of maser emission. As mentioned in the introduction, the LMC 
mass is estimated to be between 6 $\times$ 10$^{9}$ and 1.5 $\times$ 
10$^{10}$ M$_{\odot}$, giving a ratio of masses with the Galaxy of between 
approximately 30 and 80. Studies by Massey et al. (1995) and Massey (2003) 
showed the slopes of the initial mass function (IMF) of OB associations and 
massive star clusters in the LMC, SMC and Milky Way are all approximately 
equal to the Salpeter (1955) value. This suggests the metallicities of the LMC 
and SMC, which are respectively $\sim$44 and  $\sim$28 per cent of that in 
the solar neighbourhood (Massey 2003, references therein), do not affect the IMF
in the Magellanic Clouds. None the less, a deficiency of complex molecules 
resulting from these lower metallicities may result in fewer masers, if the methanol 
abundance falls below that required for maser emission. Typically for the Galaxy 
a methanol fractional abundance of 10$^{-7.5}$ $<$ X$_{\rm m}$ $<$ 10$^{-5}$ 
is required (Cragg, Sobolev \& Godfrey 2002). The recent study of M33 by 
Goldsmith et al. (2007), also concluded that the lack of detections for M33 was most 
likely due to a lower metallicity, even though the difference in metallicities between 
M33 and the Milky Way is small (M33 is $\sim$72 per cent that of the solar 
neighbourhood, Massey 2003, references therein).

\subsection{Variability}
Galactic 6668-MHz methanol masers have shown significant variability, with a 
variety of behaviours, from quasi-periodic to sporadic flares (e.g. Caswell, Vaile \& 
Ellingsen 1995; Macleod \& Gaylard 1996; Goedhart, Gaylard \& van der Walt 2004). 
When sampled at four epochs over 18 months, $\sim$25 per cent of the 6668-MHz 
methanol population are observed to show some variability (Caswell et al. 1995a). 
When considering longer periods, greater than 10 years, Ellingsen (2007) notes that
variability of the peak and integrated flux densities of the majority of masers does
not increase with the longer timescale (probably because most variations are quasi-periodic
on much smaller timescales), with only about 6 per cent showing marked variability. We
might then expect only slight variability in the LMC population. Of the four methanol sources, 
only three (N11, N105a, IRAS05011-6815) have multiple epoch observations and none 
of them show significant variation of peak flux density or spectral structure. As we are 
seeing the high luminosity tail of the LMC maser population, the stability of the masers 
observed may suggest that the LMC masers are saturated.

The completion of the MMB survey and a comparison with early extensive catalogues
such as that of Caswell et al. (1995b) will provide a more definitive characterisation of 
the long term variability properties of the Galactic 6668-MHz methanol masers (and a 
deeper survey of the LMC may provide a larger sample for comparison).

\subsection{Associations}
At the distance of the Magellanic Clouds, the typical positional uncertainties of 0.3 arcsec
correspond to $\sim$15,000 AU, so unlike high-resolution Galactic studies, the physical 
significance of association is limited. Also, of all the species, only 6668-MHz methanol 
has been seen to be exclusively associated with massive star-formation (Minier et al. 2003). 
None the less, for the star-formation regions described in section 3, we find: two incidences 
of isolated 6668-MHz methanol emission; one case of isolated 6035-MHz OH emission; four 
cases of isolated H$_{2}$O; and no isolated occurrences of ground-state OH emission. Where 
multiple transitions are observed (positions LMCm02, LMCm03, LMCm05 and LMCm11 in Table 2), the masers are seen to 
coincide within the accuracy of the observations. Due to the positional uncertainties, limited 
conclusions can be drawn on the evolutionary states the star-formation regions represent, but 
positions LMCm03, LMCm05 and LMCm11, which show ground-state OH as well as water, are likely to represent a 
later stage (Forster \& Caswell 1989).

In our own Galaxy the excited-state 6035-MHz OH maser has been observed to be commonly 
associated with 6668-MHz methanol masers (e.g. Caswell 1997, Etoka, Cohen \& Gray 2005). 
In the LMC we see one of the two 6035-MHz OH masers associated with 6668-MHz methanol 
maser emission. From the opposite view point we see one of the four 6668-MHz methanol 
masers associated with 6035-MHz OH maser emission. For this case of associated methanol and 
excited-state OH, it is interesting to note the excited-state OH is seen to be the brighter of the two, 
which from the study of Caswell (1997) may imply they signpost a region with a more developed 
H{\sc ii} region, which would also agree with the previously observed continuum emission noted 
in Section 3.

\section{Conclusions}
A complete survey of the Magellanic Clouds has been conducted as part of the Methanol Multibeam 
project and supplemented by a targeted search of known star-formation regions in the Large Magellanic 
Cloud. This has detected two new extragalactic masers: a fourth 6668-MHz methanol maser towards 
the star-forming region N160a in the LMC; and a second extragalactic 6035-MHz OH maser towards 
N157a. We have also re-observed the three previously known 6668-MHz methanol masers and the 
single 6035-MHz OH maser in the LMC. All these masers have had their positions determined with 
the ATCA and exhibit stability in spectral structure and intensity, even over the $\sim$10 year 
separation between the current and previous observations. The current lower limits on the LMC maser 
populations are demonstrated to be up to a factor of $\sim$45 smaller than the Milky Way,
with the methanol showing the largest discrepancy. Even after correcting for lower star formation rates 
(section 4.1), methanol masers in the LMC are still under-abundant by a factor of $\sim$4-5. 
This remaining disparity, in agreement with the speculation of 
Beasley et al. (1996), may be due to the lower oxygen and 
carbon abundances in the LMC. 

These observations, only made feasible by the speed provided by the 7-beam MMB receiver, have 
produced the most accurate determination yet of the luminosity distribution of extragalactic masers. 
When combined with the complete MMB Galactic catalogue, they will allow for an improved 
comparison of Galactic star-formation regions with those of lower metallicity found in the LMC. 
This will also allow us to better determine whether the LMC 6668-MHz methanol and 6035-MHz 
OH maser populations simply scale directly with the star-formation rate or if other factors are involved.

\section*{Acknowledgments}
JAG, JCox and DW-McS acknowledge the support of a Science and Technology Facilities Council (STFC) 
studentship. LQ acknowledges the support of the EU Framework 6 Marie Curie Early Stage 
Training programme under contract number MEST-CT-2005-19669 ``ESTRELA''. The Parkes Observatory 
and the Australia Telescope Compact Array are part of the Australia Telescope which is funded by the 
Commonwealth of Australia for operation as a National Facility managed by CSIRO. The authors dedicate 
this paper to the memory of R. J. Cohen.

\label{lastpage}

\end{document}